\documentclass[9pt, sigconf, screen=true,bookmarks=false]{acmart}
\AtBeginDocument{%
  }

\usepackage[utf8]{inputenc} %
\usepackage[T1]{fontenc}    %
\usepackage[table,xcdraw]{xcolor}
\usepackage{pifont}

\usepackage{algorithmicx}
\usepackage{graphicx}
\usepackage{todonotes}

\usepackage{threeparttable}
\usepackage{multirow}
\usepackage{amsmath}
\usepackage{bm}
\usepackage{float}
\usepackage{amsthm}
\usepackage{soul}
\usepackage[noend]{algpseudocode}
\usepackage{enumitem} %
\usepackage[subrefformat=parens,labelformat=parens]{subfig}
\usepackage{algorithm}
\usepackage{xspace}

\definecolor{citecolor}{RGB}{34,139,34}
\definecolor{mydarkblue}{rgb}{0,0.08,1}
\definecolor{mydarkgreen}{rgb}{0.02,0.6,0.02}
\definecolor{mydarkred}{rgb}{0.8,0.02,0.02}
\definecolor{mydarkorange}{rgb}{0.40,0.2,0.02}
\definecolor{mypurple}{RGB}{111,0,255}
\definecolor{myred}{rgb}{1.0,0.0,0.0}
\definecolor{mygold}{rgb}{0.75,0.6,0.12}
\definecolor{myblue}{rgb}{0,0.2,0.8}
\definecolor{mydarkgray}{rgb}{0.,0.2,0.2}

\definecolor{lightred}{RGB}{255,235,235}
\definecolor{lightgreen}{RGB}{235,255,235}
\definecolor{lightblue}{RGB}{235,235,255}
\definecolor{citelightblue}{RGB}{49,164,222}
\definecolor{lightcyan}{RGB}{235,255,255}
\definecolor{lightmagenta}{RGB}{255,235,255}
\definecolor{lightyellow}{RGB}{255,255,235}

\definecolor{qxkcolor}{RGB}{215,235,255}
\definecolor{softmaxcolor}{RGB}{230,235,255}
\definecolor{probxvcolor}{RGB}{255,255,235}

\definecolor{topkcolor}{RGB}{255,235,235}
\definecolor{zecolor}{RGB}{255,255,235}
\definecolor{dynacolor}{RGB}{235,255,255}

\definecolor{reviewcolor}{RGB}{0,0,200}

\newcommand{\calN}{\mathcal{N}}

\theoremstyle{plain}

\theoremstyle{definition}

\newcommand{\hpimap}{\texttt{H\textsuperscript{3}PIMAP}\xspace}
\newcommand{\calI}{\mathcal{I}}

\graphicspath{{./figs/}}
\usepackage{geometry}
\begin{document}
\vspace{-20pt}
\title{Automted Workload Mapping with 3D Heterogeneous Electronic-Photonic Processing-in-Memory Platform
}

\title{
H$^\text{3}$PiMAP: A Framework for Mapping Hybrid DNNs onto Heterogeneous 3D-Stacked Electronic-Photonic Processing-in-Memory Architectures}

\title{
H$^\text{2}$PiMAP: A Multi-Objective DNN Mapping Framework on Heterogeneous Electronic-Photonic Processing-in-Memory Architectures}

\title{
H$^\text{3}$PIMAP: A Heterogeneity-Aware Multi-Objective DNN Mapping Framework on Electronic-Photonic Processing-in-Memory Architectures
}
\author
{
Ziang Yin\textsuperscript{1}, Aashish Poonia\textsuperscript{1},
Ashish Reddy Bommana\textsuperscript{1}, Xinyu Zhao\textsuperscript{2}, Zahra Hojati\textsuperscript{1},\\ Tianlong Chen\textsuperscript{2}, Krishnendu Chakrabarty\textsuperscript{1}, Farshad Firouzi\textsuperscript{1}, Jeff Zhang\textsuperscript{1}, Jiaqi Gu\textsuperscript{1}$^\dagger$\\
\textsuperscript{1}Arizona State University, \textsuperscript{2} University of North Carolina at Chapel Hill\\
\small\textit{$\dagger$jiaqigu@asu.edu}
}
\vspace{-10pt}

\settopmatter{printacmref=false} 
\renewcommand\footnotetextcopyrightpermission[1]{} 
\pagestyle{plain} 

\begin{abstract}
\label{abstract}
The future of artificial intelligence (AI) acceleration demands a paradigm shift beyond the limitations of purely electronic or photonic architectures. 
Photonic analog computing delivers unmatched speed and parallelism but struggles with data movement, robustness, and precision. 
Electronic processing-in-memory (PIM) enables energy-efficient computing by co-locating storage and computation but suffers from endurance and reconfiguration constraints, limiting it to static weight mapping. 
Neither approach alone achieves the balance needed for adaptive, efficient AI.
To break this impasse, we study a hybrid electronic-photonic-PIM computing architecture and introduce \hpimap, a heterogeneity-aware mapping framework that seamlessly orchestrates workloads across electronic and optical tiers.
By optimizing workload partitioning through a two-stage multi-objective exploration method, \hpimap harnesses light speed for high-throughput operations and PIM efficiency for memory-bound tasks.
In system-level evaluations, \hpimap delivers a 3.32× latency reduction across language and vision models and, on large language models, achieves 77.0$\%$ lower latency with 14.6$\%$ lower energy at matched quality, outperforming homogeneous and naïve mappings strategies.
This proposed framework lays the foundation for hybrid AI accelerators, bridging the gap between electronic and photonic computation for next-generation efficiency and scalability.

\vspace{10pt}
\end{abstract}

\maketitle
\vspace{-15pt}
\section{Introduction}
\label{sec:Introduction}

Artificial Intelligence (AI), powered by deep neural networks (DNNs), now underpins a vast spectrum of applications, from computer vision and natural language processing (NLP) to medical diagnostics and time-series analytics. 
These workloads increasingly demand \textbf{sub-millisecond latency}, \textbf{high throughput}, and \textbf{stringent energy efficiency}, requirements that traditional CPU/GPU-based architectures often struggle to satisfy. 
The growing disparity between processing speed and data-movement cost, the notorious memory wall, has motivated a paradigm shift toward \textbf{domain-specific accelerators} that transcend the classical von Neumann model.

Among these emerging paradigms, \textbf{Processing-in-Memory (PIM)} and \textbf{photonic accelerators} have attracted particular attention for their complementary strengths.
PIM architectures exploit the intrinsic capability of performing arithmetic directly within or near memory arrays, drastically reducing costly data transfers between separate processing and memory units. 
By merging computation and storage, PIM architectures can significantly cut both latency and energy consumption~\cite{Sebastian2020}.
However, the landscape of PIM technologies, ranging from SRAM~\cite{zhang2017memory} to FeFET~\cite{long2019ferroelectric}, ReRAM~\cite{feinberg2018making}, and MRAM~\cite{yusuf2024domain}, presents intricate trade-offs in area, endurance, write latency, and power dissipation. 
No single technology dominates across all metrics, underscoring the need for careful technology-architectural co-design to tailor PIM solutions to specific workload demands.

Photonic accelerator systems exploit light‐based signal propagation and processing to target the heavy linear algebra that dominates DNN inference and training. 
For example, by using integrated photonic circuits to carry out matrix multiplication and accumulation, photonic computing has demonstrated sub-nanosecond inference latency and very low energy per operation.
Yet, translating these proof-of-concept results into large-scale, general-purpose DNN accelerators remains challenging. 
Key challenges include fabrication variation and non-ideal noise in optical modulators, thermal drift and crosstalk across components, limited numerical precision on high-speed optical signal modulation, and the overhead of electro-optic signal conversion.
Overcoming these limitations requires \textbf{robust calibration}, \textbf{error-tolerance}, and \textbf{co-design strategies} to fully unlock photonics’ potential in AI acceleration.
The complementary capabilities of photonics and electrical PIM offer an opportunity to leverage the best of both worlds to explore a hybrid computing paradigm.

On the application level, \textbf{the inherent heterogeneity of modern DNNs} presents both challenges and opportunities of co-design. 
Different model layers and modules exhibit varying levels of data reuse, computational intensity, and sensitivity. 
For instance, mapping highly dynamic attention layers in Transformers onto non-volatile PIM arrays can lead to rapid device wear-out~\cite{Chi2016PRIME}, while photonics might excel for dense, static matrix multiplications but is less suited for depthwise or data-transformation-heavy operations.
These diverse workload characteristics of modern DNN models imply a perfect match with heterogeneous electronic-photonic-PIM hardware architecture, but demand \textbf{heterogeneity-aware mapping strategies} that intelligently allocate tasks to the most suitable accelerator type, maximizing efficiency without compromising accuracy or device lifetime.

To enable efficient workload mapping onto next-generation electronic-photonic-PIM accelerators, we propose \hpimap, a heterogeneity-aware multi-objective mapping scheme, an effective two-stage mapping flow that jointly optimizes energy, latency, and robustness:

\indent\textbf{1.} \textbf{Pareto-optimal exploration stage} identifies optimal energy-latency configurations across hybrid PIM-photonic platforms using integrated architectural and interconnect simulation.

\indent\textbf{2.} \textbf{Sensitivity-aware remapping stage} refines the initial mapping to preserve model accuracy under hardware non-idealities, ensuring resilience against device variability and quantization effects.

Through this integrated design and mapping approach, \hpimap enables efficient cooperation between electronic and photonic accelerators for hybrid AI workloads.
Our key contributions are as follows:
\begin{itemize}[leftmargin=*]
\setlength{\itemindent}{0.5em}
\vspace{-8pt}
    \item \textbf{Electronic-Photonic-PIM Accelerator Design}: 
    We introduce a novel hybrid accelerator that combines PIM hardware with photonics and achieves superior inference power, speed, and accuracy over homogeneous systems.
    \item \textbf{Electronic-Photonic-PIM Accelerator Evaluation Infrastructure}: 
    We create the first system modeling and optimization framework integrating SoTA 
    simulators for PIM, photonic AI hardware, 
    
    and network-on-chips, enabling automated design exploration.
    \item \textbf{Heterogeneity-aware Mapping Framework}: 
    We introduce a two-stage workload mapping framework that considers a workload's variability and target hardware's heterogeneity. This ensures rapid mapping for space exploration for optimal resource utilization, system performance, and inference accuracy.
    \item \textbf{Comprehensive Performance Evaluation}:
    Across diverse NLP and vision benchmarks, \hpimap  delivers 3.32$\times$ lower latency than homogeneous systems and naïve mappings; on large language models (LLMs), the gains are even stronger, 77.0$\%$ lower latency and 14.6$\%$ lower energy under matched-quality constraints.
\end{itemize}

\section{Related Work}
\label{sec:prelim}

\vspace{-2pt}
We first survey DNN accelerators built entirely from a single technology---SRAM-, ReRAM-based PIM, or photonics, and compare their characteristics in Table~\ref{tab:FeatureCompare}.

\textbf{PIM Accelerators: } Numerous PIM~\cite{zhang2017memory, long2019ferroelectric, feinberg2018making, yusuf2024domain, shafiee2016isaac, Chi2016PRIME, 10493861} designs push matrix multiplication directly into memory arrays, cutting data movement but trading off write latency, endurance, or noise. Recent surveys~\cite{joardar2020accured} indicate that no single device dominates all metrics, motivating the adoption of hybrid solutions. 

\textbf{Photonic Tensor Cores: }  Photonic tensor cores~\cite{NP_Nature2021_Xu, NP_NATURE2017_Shen, NP_SciRep2017_Tait, NP_ACS2022_Feng, NP_HPCA2024_Zhu} deliver sub-nanosecond reconfiguration, passive device matrix computation, and >800 TOPS throughput. Their combined power-performance-area (PPA), therefore, surpasses any known digital accelerator. Nevertheless, thermal drift and ADC/DAC overheads still cap full-system efficiency, and faithfully modeling these non-idealities remains an open challenge.

\textbf{Hybrid \& Heterogeneous Accelerators: }  Prior work couples SRAM with ReRAM for MSB/LSB splitting or combines GPUs with PIM arrays, yet largely optimizes a single objective or ignores photonics altogether.~\cite{rashed2021hybrid, bhattacharjee2023hyde, Luo2024H3DTransformerAH} To the best of our knowledge, no prior work integrates photonics with PIM devices for DNN acceleration while simultaneously tackling layer-to-device assignment under joint latency, energy, and accuracy constraints, a gap that motivates our \hpimap framework for efficiently partitioning and mapping DNN workloads onto heterogeneous electronic–photonic–PIM accelerators.

\section{Proposed \hpimap Framework on Electronic-Photonic-PIM Architecture}
\label{sec:method}
In this section, we present a heterogeneous electronic-photonic-PIM architecture, analyze its key characteristics and device noise, and describe \hpimap framework that exploits these features for efficient, robust DNN mapping.
\subsection{Architecture Overview and Analysis}
\label{sec:ArchAnalysis}
\begin{figure}
    \centering
    \includegraphics[width=1\columnwidth]{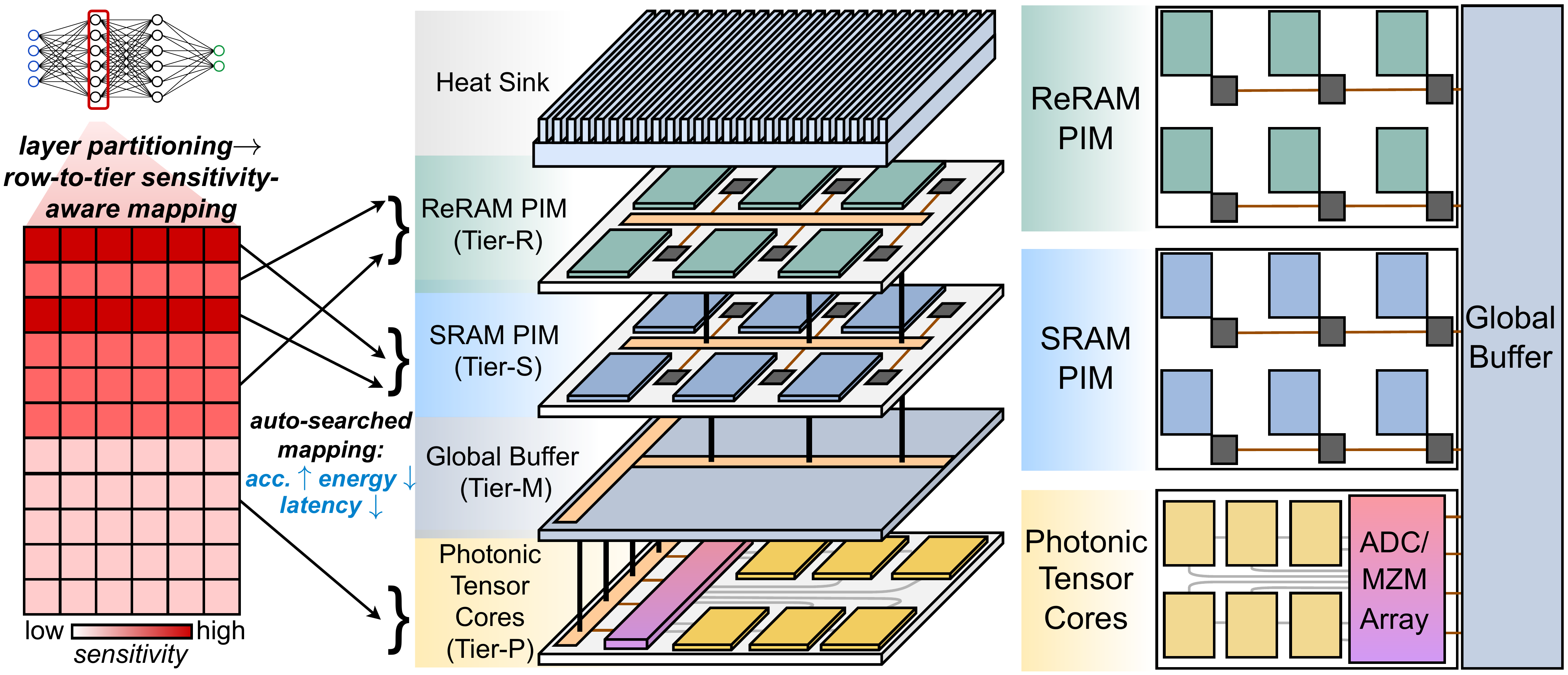}
    \caption{2D/3D heterogeneous electronic-photonic-PIM architecture with ReRAM, SRAM, and photonics.}
    \label{fig:combine_arch}
    \vspace{-8pt}
\end{figure}

As a concrete case, we consider a four-tier 3D-stacked accelerator architecture: a ReRAM-PIM layer (R), an SRAM-PIM layer (S), an integrated photonic layer (P), and a global-buffer layer (M). Each tier hosts many identical tiles built in a single technology and silicon through vias (TSVs) ferry data between tiers. The shared buffer in tier M holds photonic weights and activations for all three compute layers.

As an important motivation, we answer a critical question: Why bother fusing R, S, and P into a heterogeneous architecture?

\noindent\textbf{Feature Analysis of 3 Technologies} -- Table~\ref{tab:FeatureCompare} highlights the three technologies' speed, energy, and precision in complementary ways, compelling a heterogeneity-aware mapper.
\begin{table}
\caption{Comprehensive comparison of multi-tile accelerators with SRAM, ReRAM, and photonics. }
\resizebox{\columnwidth}{!}{
\begin{tabular}{c|c|c|c}
\hline
Property & SRAM 22nm                            & ReRAM 32nm                            & Photonics  \\ \hline 
Resolution & 1-bit $\times$ 8 cells=8-bit                             & 2-bit $\times$ 4 cells = 8-bit                             & 4$\sim$6-bit  \\ \hline
Tile Size & 256 crossbars, 128$\times$128                              & 64 crossbars, 128$\times$128                            & 2 cores, 14$\times$14  \\ \hline
ADC/tile & 256 SARADC 7-Bit & 64 SARADC 8-Bit & 392 SARADC 8-Bit\\ \hline
Cell Area & $<1\mu m^2$                              & $<1\mu m^2$                            & $>1000\mu m^2$  \\ \hline
Arch Size & 100 Tiles & 100 Tiles & 2 Tiles\\ \hline
Program latency & $\sim$ 1 ns                              & $\sim$ 100 ns                            & $\sim$ 100 ps  \\ \hline
Static power &   Medium                            & Low                            & High  \\ \hline
Clock & 100 MHz                              & 100 MHz                            & 3 GHz \\ \hline

\end{tabular}
}

\label{tab:FeatureCompare}
\vspace{-10pt}
\end{table}

\noindent\ding{202}~\textbf{Speed}: Photonic cores race through math, 1–10 TOPS with only a handful of tensor units, so they shine on compute-bound layers (convs, self-attention). ReRAM PIM is slower, throttled by ADC sampling, yet fantastically energy-wise for memory-bound matrix multiplies. SRAM PIM lands between, giving balanced throughput for medium workloads.

\noindent\ding{203}~\textbf{Static vs. dynamic weights:} ReRAM and SRAM favor weight-static workloads, weights stay resident to keep device endurance (e.g., ReRAM). Dynamic layers that rewrite both operands (e.g., Q $\times$ K in Transformers) suit photonic cores better.

\noindent\ding{203}~\textbf{Precision and noise:} SRAM/ReRAM handle 8-bit MACs but run hot, and 2-bit ReRAM cells drift. Photonic cores perform one-shot light-speed multiplies yet are limited to <6 bits and suffer thermal crosstalk and fabrication noise—calibration is essential.

\vspace{-4pt}
\subsection{Dataflow and Interconnect Design}
To match the distinct properties and operational demands of both PIM and photonics, we customize a dataflow to partition matrix multiplication in each DNN layer among three computing tiers (R/S/P). 
As each PIM tile communicates solely with the global buffer via TSVs and routers, data transfer remains one-dimensional. 
This approach eliminates inter-tile communication overhead. 
To further optimize communication efficiency, TSV connections to the PIM tiers are positioned midway between PIM tiles. 
This placement effectively halves the average communication distance relative to a 2D network-on-chip. 
Additionally, we add a dedicated TSV link to the photonic tier, accommodating its substantial bandwidth requirements. 
Our following introduced mapping algorithm and system evaluations are based on the above dataflow and interconnect designs.

\vspace{-5pt}
\subsection{Non-ideal Hardware Noise Modeling}
Analog ReRAM and photonic tiers suffer thermal, shot, and phase noise that threaten accuracy, so we apply established per-tier noise models to steer a robustness-aware mapper.
\begin{itemize}[leftmargin=*]
\setlength{\itemindent}{0.5em}
    \item \textbf{PIM}: 
    SRAM’s high tolerance and digital logic render it immune to thermal noise~\cite{10493861}, while thermally sensitive ReRAM faces dominant thermal and shot noise~\cite{joardar2020accured}, formulated as:
    \begin{equation}
       \small
        \begin{aligned}
            \Delta G_{\text{thermal}} &= \mathcal{N}\left( 0, \sqrt{4G \cdot \text{Freq} \cdot K_B \cdot T/V} \right), \\
            \Delta G_{\text{shot}} &= \mathcal{N}\left( 0, \sqrt{ 2G \cdot \text{Freq} \cdot q/V} \right),
        \end{aligned}
    \end{equation}
    where \(G\),~\(V\),~\(K_b\),~\(T\), and \( Freq \) are conductance, voltage, Boltzmann constant, temperature, and frequency, respectively.
    \vspace{2pt}
    \item \textbf{Photonics}:
    To model photonic noise, we adopt the TeMPO architecture~\cite{zhang2024tempo} and its device-level measurements. Following TeMPO, we model noise as a relative Gaussian perturbation on each matrix input, i.e., $\widetilde{X}_q=X_q + \Delta X$, where $\Delta X\sim \calN(0, (\sigma|X_q|)^2)$. We fix \(\sigma = 0.0031\) for the photonic tier.

\end{itemize}
\vspace{-5pt}
\subsection{Mapping Problem Formulation}
\begin{figure*}
    \centering
    \includegraphics[width=0.98\textwidth]{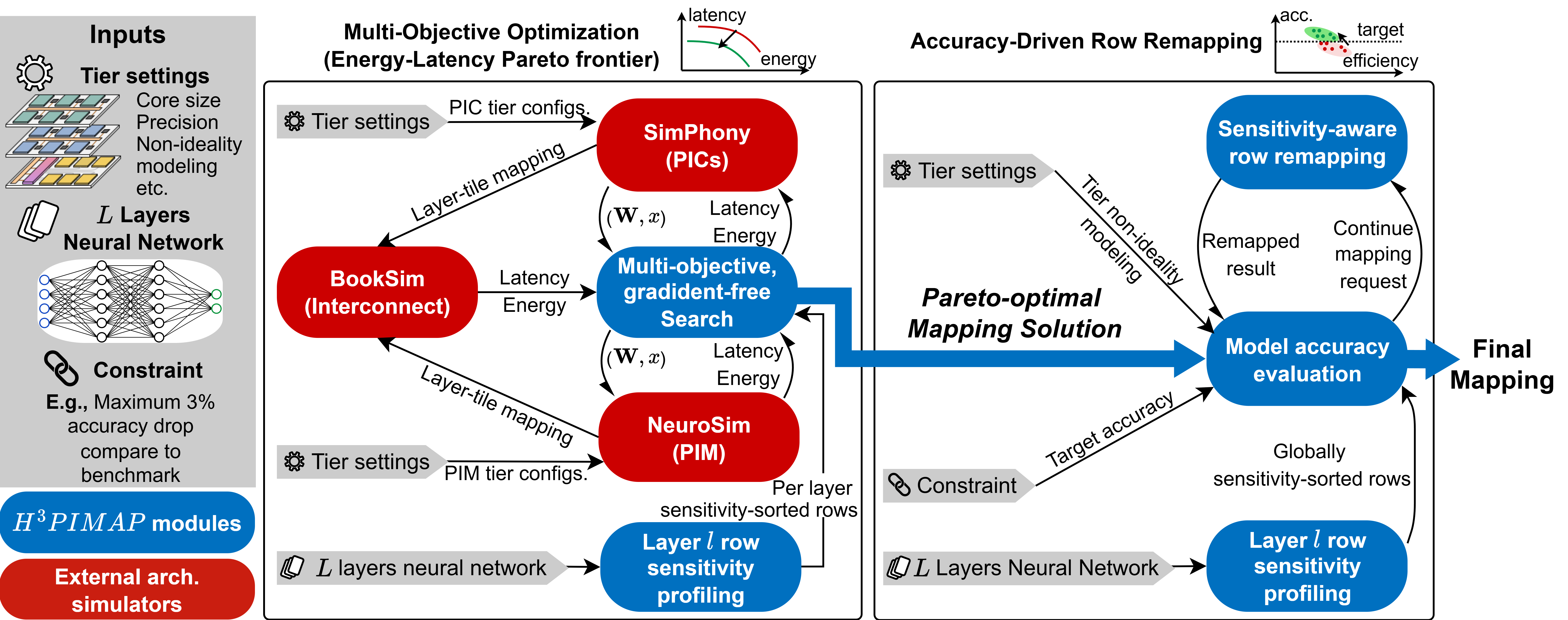}
    \caption{Overview of our heterogeneous layer-to-hardware mapping flow. Stage 1 explores the Pareto-optimal mappings in the latency-energy space. Stage 2 adjusts mapping to trade efficiency for higher accuracy until the target accuracy is met.}
    \label{fig:Flowchart}
\end{figure*}

Given an $L$-layer DNN with sequential layer execution, our goal is to obtain an optimal matrix row-to-tier mapping $\bm{\alpha_l}=(\alpha_{l, 1},\alpha_{l, 2},\cdots,\alpha_{l, n})$ individually for each layer on our heterogeneous architecture.
Each matrix row is assigned to one computing tier.
We use $\alpha_i$ to denote the percentage of matrix rows mapped to tier $i$.
The objective of mapping is to minimize the overall model inference energy ($\bm{E}$) and latency ($\mathbf{LAT}$) under several specific constraints.
This can be formulated as the following multi-objective optimization (MOO) problem:
\begin{equation}
    \small
    \vspace{-3pt}
    \label{eq:ProblemFormulation}
    \begin{aligned}
        \min_{\aleph} F(\aleph)=&(\mathbf{LAT}(\aleph), \bm{E}(\aleph)),\\ 
        \aleph =& (\bm{\alpha_1},~\bm{\alpha_2},~...,~\bm{\alpha_L})\\
        \bm{\alpha_l} =&(\calI_{l, 1}, \calI_{l, 2}, ..., \calI_{l, R_l}), \calI\in[n]\\
        \bm{\alpha_{l,i}} =& \{\calI\in\bm{\alpha_l}|\calI=i\}, \forall i\in[n]\\
        \mathbf{LAT}(\aleph)=&\sum^{L}_{l=1}\big(\max_{i\in[n]}~LAT_{i}(\bm{\alpha_{l,i}})\big)\\
        \mathbf{E}(\aleph) =&\sum^L_{l=1}\sum^n_{i=1} E_i(\bm{\alpha_{l,i}})\\
        \text{s.t.}~~
        \bm{\alpha_{l,i}}&\in\emptyset~\text{if op $l$} \text{ is not supported by tier $i$}\\
        Acc_0 & - Acc(\aleph) \leq \tau
    \end{aligned}
\end{equation}
The final configuration $\aleph=~(\bm{\alpha_1},~\bm{\alpha_2},~...,~\bm{\alpha_L})$ specifies how each layer’s weight rows are assigned across $n$ tiers.
Total latency is the sum of layer latencies, each gated by its slowest tier.
Total energy is the sum of per-layer energies.
As discussed in Section~\ref{sec:ArchAnalysis}\ding{203}, any operation that a given tier cannot support will not be mapped to that tier.
Because device noise and limited precision can degrade task performance, our noise-aware mapping framework requires the post-mapping accuracy $Acc(\aleph)$ to remain within $\tau$ discrepancy below the original pretrained accuracy $Acc_0$.

\noindent\textbf{Search Space Analysis}:~
The complexity of the problem can be illustrated with the following example. 
For an $L$-layer DNN with $R$ neurons (weight rows) per layer mapped onto $n$ tiers, there are total $\calI(n^{RL})$
possible mappings.
For example, a popular language model Pythia-70M has an average of 2048 neurons per layer and 6 layers in total. 
With 3 tiers to map, the total search space is 3$^{12288}$.
Besides the vast search space, the non-trivial cost of evaluating the accuracy and hardware efficiency of each solution poses significant challenges to the MOO task.
To efficiently explore the design space of this constrained multi-objective optimization problem, we propose \hpimap, a two-stage search and optimization procedure as shown in Fig.~\ref{fig:Flowchart}. 
We decouple the accuracy and efficiency optimization in two stages and prune the search space with sensitivity-aware heuristics.
The first stage rapidly explores the energy-latency space to obtain Pareto-optimal mapping candidates.
Then, the second stage performs accuracy-driven row remapping to strategically calibrate the row assignment to satisfy the accuracy constraints.

\subsubsection{Stage 1: Latency-Energy Pareto Optimization (PO)}
\label{sec:ST1}
In the first stage of \hpimap, we perform a multi-objective search that focuses on exploring the Pareto front of energy–latency trade-offs.
Since only the number of matrix rows mapped to a tier impacts the system inference latency and energy, not specific row indices, we are allowed to significantly reduce the search cost in the exponential mapping space.
The pruned search space now reduces from $n^{RL}$ combinations to a much smaller space with only $(\binom{R+n-1}{n-1})^L$ combinations.
As an example, for Pythia-70M, the search space is reduced from $3^{12288}$ to 8.6$\times$10$^{37}$.

We employ a tailored evolutionary algorithm NSGA-II~\cite{deb2002fast} to solve stage 1, which queries the tier-specific architecture simulators for the fitness of a particular mapping. 
The algorithm of stage 1 is detailed in Alg.~\ref{alg:NSGA-II}.
The simplified energy-latency Pareto optimization is
\begin{equation}
    \small
    \label{eq:Stage1Formulation}
    \begin{aligned}
        \min_{\aleph} F(\aleph)=&(\mathbf{LAT}(\aleph), \bm{E}(\aleph)),\\ 
        \aleph = (\bm{\alpha_1},,~...,&~\bm{\alpha_L}), 
        \bm{\alpha_l} =(\alpha_{l, 1}, ..., \alpha_{l, n}),\alpha\in\{0,\cdots,n\}\\
        \mathbf{LAT}(\aleph)\!=\!\sum^{L}_{l=1}&\big(\max_{i\in[n]}LAT_{i}(\alpha_{l, i})\big),\mathbf{E}(\aleph) \!=\!\sum^L_{l=1}\sum^n_{i=1} E_i(\alpha_{l, i})\\
        \text{s.t.}~~
        \sum^n_{i=1}& \alpha_{l,i} = R_l;~\alpha_{l, i} \in\mathbb{Z},~\forall l\\
        \alpha_{l,i}& \;=\; 0~\text{if op $l$} \text{ is not supported by tier $i$}\\
        Acc_0 & - Acc(\aleph) \leq \tau
    \end{aligned}
\end{equation}
\begin{algorithm}[t]
\small
\caption{Energy-Latency Pareto Optimization (PO)}
\label{alg:NSGA-II}
\begin{algorithmic}[1]
    \State \textbf{Input:} $N$: Population size, $G$: maximum generations,\ $p_{\mathrm{co}}$: crossover rate, $p_{\mu}$: mutation rate
    \State \textbf{Output:} $S$: Pareto-front solution set 

    \State \textbf{Initialize} population $S$ of size $N$ with random tier-assignment percentages
    \State \textbf{Evaluate} each solution $\aleph \in S$ by calling simulators (e.g., NeuroSim~\cite{peng2019dnn+}, SimPhony~\cite{yin2024simphonydevicecircuitarchitecturecrosslayermodeling}, BookSim~\cite{jiang2013detailed}) to measure \textbf{LAT} and \textbf{E}

    \For{$t = 0$ \textbf{to} $G - 1$}
        \State \textbf{Non-Dominated Sorting} on $S$ to obtain fronts $F_1, F_2, \dots, F_k$
        \State \textbf{Compute Crowding Distance} within each front
        \State \textbf{Select Parents} from $S$ (e.g., via tournament) based on rank (front) and crowding distance
        \State \textbf{Generate Offspring} using Crossover($p_{\mathrm{co}}$) and Mutation($p_{\mu}$)
        \State $Offspring \gets \{\, ch \in Offspring \mid \text{Feasible}(ch) \}$ \text{(e.g., operation support)} 
        \State \textbf{Evaluate Offspring}: for each child, call simulators to obtain \textbf{LAT} and \textbf{E}
        \State \textbf{Combine} $S$ and Offspring $\rightarrow R$
        \State \textbf{Non-Dominated Sorting} on $R$ to get $F_1, F_2, \dots, F_m$
        \State $S \gets \emptyset$; \quad $\ell \gets 1$
        \While{$|S| + |F_\ell| \leq N$}
            \State $S \gets S \cup F_\ell$
            \State $\ell \gets \ell + 1$
        \EndWhile
        \If{$|S| < N$}
            \State \textbf{Compute Crowding Distance} on $F_\ell$
            \State \textbf{Sort} $F_\ell$ by descending crowding distance
            \State Add the top $(N - |S|)$ individuals from $F_\ell$ to $S$
        \EndIf
    \EndFor

    \State \textbf{Return} $S$ \;  \Comment{Final Pareto front}
\end{algorithmic}
\end{algorithm}

Once we obtain the Pareto-front populations with high efficiency and speed, we examine whether the best-accuracy solution among them meets the accuracy constraint.
If such a solution exists, the best-accuracy one will be the final solution. If not, we will pass the best-accuracy one to the second stage to adjust the mapping until the accuracy target is met.  
Due to device-specific bit precision and noise, we strategically use row-wise sensitivity to guide the assignment. Specifically, we compute the sensitivity of each row \(\mathbf{W}_{l,r}\) using a second-order Taylor expansion with Gaussian perturbations:
\begin{equation}
    \small
    \label{eq:LayerAssignment}
    \begin{aligned}
        S_{W_{l, r}} \!\!=~\!\!\mathcal{L} \!-\! \mathcal{L}_{0}\!\approx\! (\nabla_{\mathbf{W}}\mathcal{L})^\top \!\!\Delta \mathbf{W}_{l, r} \!\!+\!\! \frac{1}{2} ({\nabla^2_{\mathbf{W}} \mathcal{L}})^\top\!\! \Delta \mathbf{W}^2_{l, r}
    \end{aligned}
\end{equation}
where $\mathcal{L}$ is the loss function,  $\nabla_{\mathbf{W}_{l,r}}$ is the gradient of the loss with respect to one row $r$ in layer $l$, the hessian matrix approximated from its diagonal entirety $\nabla_{\mathbf{W}_{l, r}}^{2}$, and $\Delta \mathbf{W}_{l, r}$ represents perturbations in row $r$ of the weight matrix.
A sorted tier from \textbf{best to worst model performance} $T = (t_1, t_2, ..., t_n)$ can be acquired by measuring the precision of each tier on the same workload.
Next, using the row-wise sensitivity information, we perform a sorted assignment between the rows and tiers, i.e., mapping the \textbf{most sensitive rows to the most accurate tiers}.

Finally, we evaluate the configuration to ensure the overall accuracy meets the required threshold. If so, we have a valid mapping configuration that satisfies both performance requirements and accuracy. Otherwise, we choose the mapping with the highest model performance ($\aleph_{best~perf}$) and proceed to the second stage, monotonic optimization, to further improve accuracy.
\vspace{-2pt}
\begin{algorithm}[t]
\caption{Accuracy-Driven Row Remap (RR)}
\label{alg:MonotonicMapping}
\begin{algorithmic}[1]
\small

\State \textbf{Input:} $\aleph_{\mathrm{best}\,\mathrm{perf}}$: A Pareto-front mapping configuration, \(\tau\): Accuracy-degradation threshold, 
\(\delta\): Step size for shifting assignments, \(T = [t_1, t_2, \dots, t_k]\): Tiers sorted from \emph{best} (least noise) to \emph{worst} (most noise)

\State \textbf{Output:} \(\aleph^*\): Final weight mapping such that \( \mathrm{Acc}(\aleph^*) - \mathrm{Acc}_0 \le \tau.\)

\State \(\aleph \gets \aleph_{\mathrm{best}\,\mathrm{perf}}\) 
    \Comment{Initialize from best-performance configuration}
\State \(\mathrm{AccCurrent} \gets \mathrm{Evaluate}(\aleph)\)
    \Comment{Evaluate the mapping config from Stage 1}

\While{\(\mathrm{Acc}_0 - \mathrm{AccCurrent} > \tau\)} 
    \Comment{We still need to improve accuracy}
    \State \(\mathrm{worstTier} \gets \text{findWorstTierWithUsage}(l, \aleph)\)
        \Comment{e.g., from end of \(T\) that still has \(\alpha_{l,\mathrm{worst}} > 0\)}
    \State \(\mathrm{bestTier} \gets \text{findBestTierWithUsage}(l, \aleph)\)
        \Comment{e.g., from start of \(T\) that \(\alpha_{l,\mathrm{best}} < 1\)}
    \If{\(\mathrm{worstTier} = \varnothing\) \textbf{or} \(\mathrm{bestTier} = \varnothing\)}
        \Comment{No more shifting possible}
        \State \textbf{break while}
    \EndIf

    \State \(\Delta = \min\bigl(\bm{\alpha}_{l,\mathrm{worstTier}},\, \delta\bigr)\)
        \Comment{Shift up to \(\delta\), but cannot exceed what is allocated.}
    \State \(\bm{\alpha}_{l,\mathrm{worstTier}} \gets \bm{\alpha}_{l,\mathrm{worstTier}} - \Delta\)
    \State \(\bm{\alpha}_{l,\mathrm{bestTier}} \gets \bm{\alpha}_{l,\mathrm{bestTier}} + \Delta\)

    \State \(\mathrm{AccCurrent} \gets \mathrm{Evaluate}(\aleph)\)
        \Comment{Re-evaluate after shifting}
    \If{\(\mathrm{Acc}_0 - \mathrm{AccCurrent} \le \tau\)}
        \State \textbf{break while} 
            \Comment{Accuracy requirement satisfied}
    \EndIf
\EndWhile

\end{algorithmic}
\end{algorithm}

\subsubsection{Stage 2: Acc-Driven Row-Remap (RR)}
In the accuracy optimization stage, we iteratively reassign weight matrix rows from the most noise-prone tier to the tier offering higher accuracy to ensure the process converges to the required accuracy.
Although the initial weight mapping $\aleph_{best~perf}$ (acquired in Stage 1) provides a per-layer distribution, we convert it into a global distribution across all tiers for the entire model. 

\subsection{System Feasibility of \hpimap}
The \hpimap architecture adopts a four-tier 3D-stacked design, with each tier fabricated separately and integrated via standard packaging solutions.
This modular approach avoids the yield issues of monolithic integration and is compatible with existing foundry technologies.
Its feasibility is supported by increasing academic and industrial demonstrations.
Prototypes have stacked SRAM on PICs for near-memory photonic control~\cite{Pintus2025MagnetoOptics}, and bonded photonic-electronic test chips show compatibility with standard 3D stacks without exceeding power, thermal, or yield constraints~\cite{Hsia2021COUPE}.
Researchers are exploring fully optical PIM fabrics that eliminate electronic data movement~\cite{Rios2019PhotonicIMC}, and industry efforts, such as Lightmatter’s co-packaged optical compute planes with CMOS memory and logic~\cite{long2019ferroelectric}, further validate the practicality of heterogeneous 3D electronic-photonic-PIM integration explored in this emerging study.

\begin{table}[htp]
\centering
\caption{Dataset specifications.}
\resizebox{1\columnwidth}{!}{
\begin{tabular}{|c|c|c|c|c|}
\hline
Datasets        & Dataset Size & Training Set  & Test Set     & Benchmark                                                                   \\ \hline
TinyStories     & 1.92GB       & 2.12M rows    & 22K rows     & \begin{tabular}[c]{@{}c@{}}20.329 (PPL 70M)\\ 14.334 (PPL 2.8B)\end{tabular} \\ \hline
Military Assets & 4.19 GB      & 21978 samples & 1396 samples & 0.8972 (Acc)                                                                \\ \hline
Chest X-Ray     & 1.24 GB      & 5216 samples  & 624 samples  & 0.9533 (Acc)                                                                \\ \hline
\end{tabular}
}
\label{tab:Datasets}

\end{table}

\begin{table}[htp]
\centering
\caption{Model settings and their layer counts.}
\resizebox{1\columnwidth}{!}{
\begin{tabular}{|c|c|c|c|c|c|c|}
\hline
Models      & Params & \begin{tabular}[c]{@{}c@{}}Num\\ Layers\end{tabular} & \begin{tabular}[c]{@{}c@{}}Num\\ Linear\end{tabular} & \begin{tabular}[c]{@{}c@{}}Num\\ Conv2d\end{tabular} & \begin{tabular}[c]{@{}c@{}}Num\\ Attention\end{tabular} & \begin{tabular}[c]{@{}c@{}}Num\\ Matmul\end{tabular} \\ \hline
MobileViT-S & 5.6 M  & 69                                                   & 37                                                   & 32                                                   & 9                                                       & 18                                                   \\ \hline
Pythia-70M      & 70 M   & 24                                                   & 24                                                   & 0                                                    & 6                                                       & 12                                                   \\ \hline
Pythia-2.8B      & 2.8 B   & 128                                                  & 128                                                  & 0                                                    & 32                                                      & 64                                                   \\ \hline
\end{tabular}
}
\label{tab:ModelUsed}
\end{table}

\section{Evaluation Results}
\label{sec:Exp}
\subsection{Evaluation Setup}
\label{sec:Setup}

\begin{figure}[htp]
    \centering
    \includegraphics[width=0.5\columnwidth]{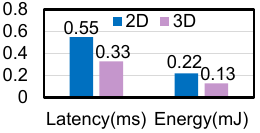}
    \caption{Communication cost b/w 2 Conv2D layers w/ input size [8, 3, 32, 32] and [8, 16, 32, 32] in a 10$\times$10 PIM mesh.}
    \label{fig:Interconnect2D3D}
\end{figure}

Our experiments use the Pythia-70M~\cite{biderman2023pythia} trained on the language task TinyStories dataset~\cite{eldan2023tinystories} as baseline for analysis. 
We evaluate the MobileViT-S vision model~\cite{mehta2021mobilevit} on two distinct datasets: a military asset~\cite{rawsi18_2023} and a medical chest X-ray dataset~\cite{jaeger2014two} to prove our framework's versatility. All models and datasets are shown in Table~\ref{tab:Datasets} and~\ref{tab:ModelUsed}.

The models we evaluate retain their original architectures, modified by learned quantization~\cite{esser2019learned}.
We first train all models from scratch in an 8-8-8-bit (input–weight–output) configuration, then fine-tune a 6-6-8-bit variant from the 8-bit checkpoint.
This fine-tuning step helps maintain a smooth distribution when moving to lower bit precision, accommodating tiers with constrained precision. 
Note that this fine-tuned 6-6-8-bit checkpoint is used whenever a workload is evaluated on a configuration that assigns any portion to the photonic tier. All experiments are performed on a Linux OS  server with NVIDIA A6000 GPUs and AMD EPYC 7763 64-core processors.

\subsection{Results}\label{sec:Results}
\subsubsection {Network-on-Chip Interconnect Cost}

Using BookSim NoC simulator~\cite{jiang2013detailed}, we simulate the performance of \textbf{2.5D} and \textbf{3D} interconnect topologies with inputs of size [8, 3, 32, 32] between two Conv2D layers, shown in Fig.~\ref{fig:Interconnect2D3D}.
The experiment shows an improvement of \textbf{40\%} and \textbf{41\%} in latency and energy cost, respectively.

\begin{figure}
    \centering
    \includegraphics[width=0.65\columnwidth]{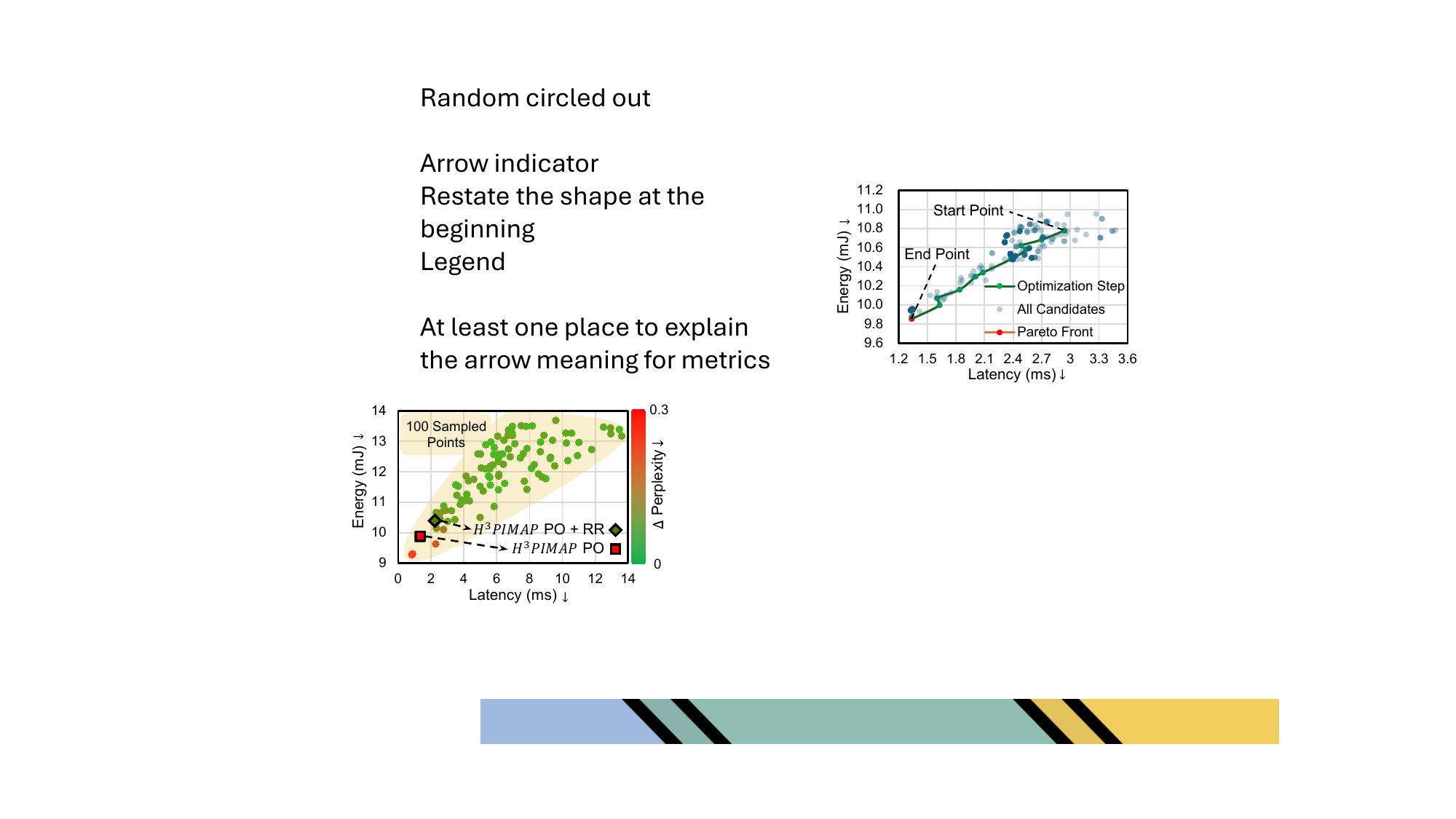}
    \caption{Energy/latency improves during stage 1 search.}
    \vspace{-10pt}
    \label{fig:NSGAPareto}
\end{figure}

\begin{figure}[htp]
    \centering
    \includegraphics[width=0.99\columnwidth]{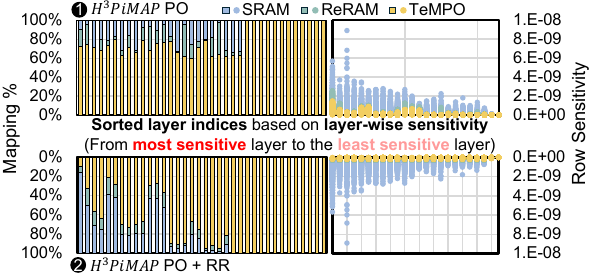}
    \caption{The Pythia-70M's layer-wise workload distribution and row-assignment among three devices. \ding{202} upper two figures show the workload distribution of \hpimap Pareto optimization (PO), and \ding{203} lower two shows that of \hpimap Pareto optimization (PO) + row remapping (RR).}
    \label{fig:BeforeMono}
\end{figure}

\subsubsection{Pareto Optimization (PO)}
We initialize the population and let the NSGA-II ~\cite{deb2002fast} algorithm seek Pareto-optimal mappings for Pythia-70M. Figure~\ref{fig:NSGAPareto} charts energy–latency co-optimization during the evolutionary run, while Fig.~\ref{fig:BeforeMono} \ding{202} depicts layer-wise workload across three tiers (left) and row-wise mapping per layer (right) after PO.
\subsubsection{Accuracy-Driven Row-Remapping (RR)}
\begin{figure}
    \centering
    \vspace{-12pt}
    \includegraphics[width=0.85\columnwidth]{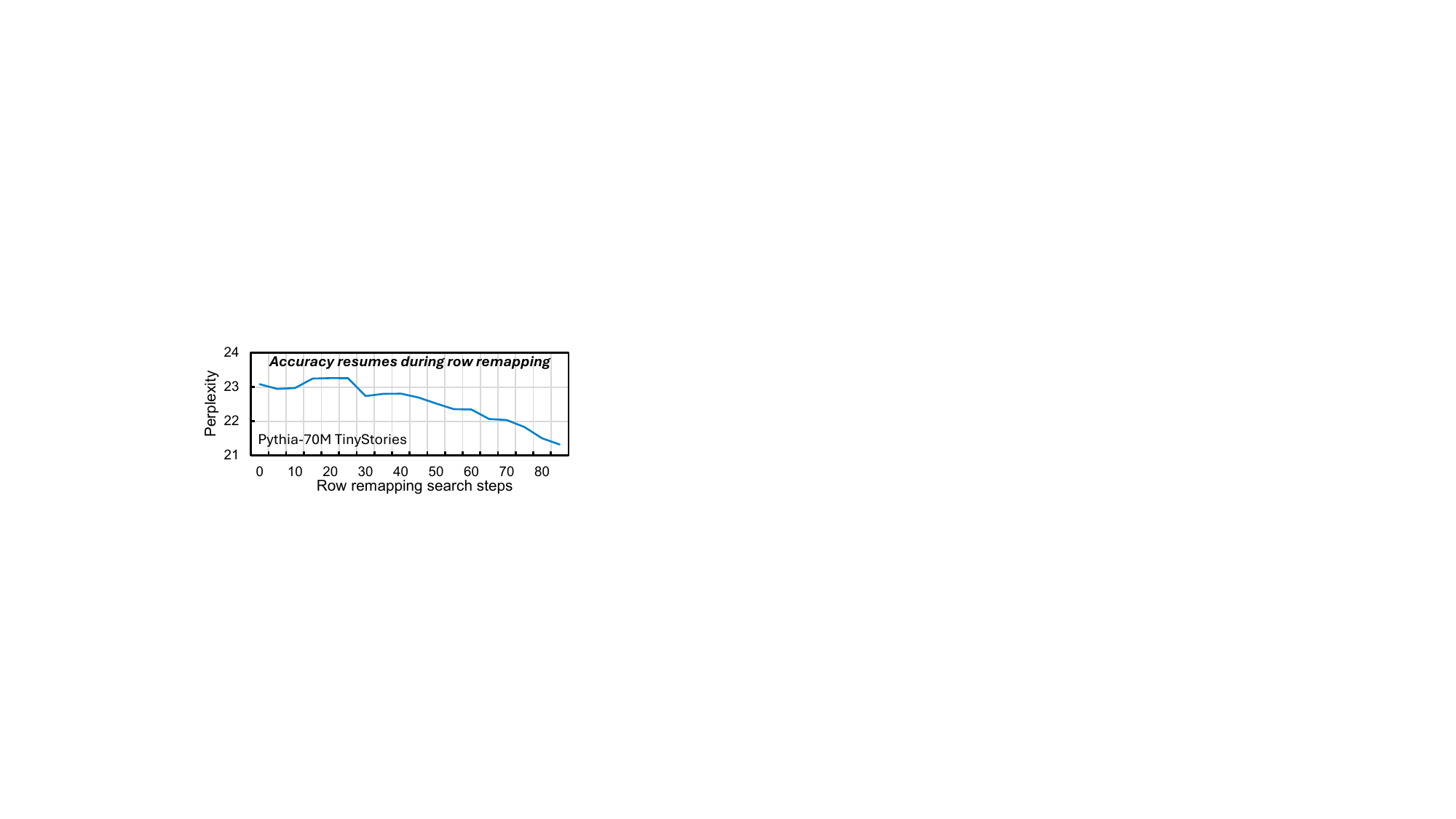}
    \caption{The PPL search improves during second-stage row remapping on Pythia-70M TinyStories.}
    \label{fig:MONOOPT}
\end{figure}

Once we obtain the set of Pareto-optimal solutions, there are inevitably multiple candidates that balance our objectives in different ways. To narrow this down, we explicitly evaluate the accuracy for each of these candidates and select the one that yields the best final accuracy. As a preliminary step, we calculated each row’s sensitivity using a Taylor expansion approach using Eq.~\eqref{eq:LayerAssignment}. Figure~\ref{fig:BeforeMono}~right side shows the row-wise sensitivity distribution of each layer, sorted from the most sensitive layer to the least sensitive one.
Leveraging these sensitivity insights, we then gradually shift a portion of the photonic chip’s workload towards SRAM. Since SRAMs are not affected by low-precision and noise-related issues, this rebalancing can help retain the model's accuracy. 
In Fig.~\ref{fig:MONOOPT}, the PPL steadily improves as the optimization progresses. We impose a 1.0 tolerance relative to the noise-free PPL for RR. Comparing the final workload assignment after RR (Fig.~\ref{fig:BeforeMono}~\ding{203}) with the one in PO (Fig.~\ref{fig:BeforeMono}~\ding{202}), we observe a noticeable shift in workload from the Photonic tier to SRAM.

\begin{table}
\centering
\vspace{-10pt}
\caption{Mapping strategy comparison tested on Pythia-70M/TinyStories. PIM and photonics are designed on ISAAC~\cite{shafiee2016isaac} and
TeMPO~\cite{zhang2024tempo} architectures, respectively.}
\resizebox{1\columnwidth}{!}{
\begin{tabular}{c|c|c|c|c|c}
\hline
Strategy                  & Latency(ms)$\downarrow$ & Energy(mJ)$\downarrow$ & \begin{tabular}[c]{@{}c@{}}Precision\\ (In-W-Out)\end{tabular} & Perplexity$\downarrow$      & LEP Score$\downarrow$       \\ \hline
100\% SRAM                & 10.21         & 13.79        & 8-8-8                                                          & 20.329
          & 1.673      \\ \hline
100\% ReRAM               & 14.73         & 13.44        & 8-8-8                                                          & 20.340
          & 1.931      \\ \hline
100\% TeMPO               & 0.91          & 8.92         & 6-6-8                                                          & 23.839
          & 1          \\ \hline
Equal Distribution        & 4.90         & 12.02       & Mixed                                                          & 22.413
          & 1.519          \\ \hline
\rowcolor[HTML]{C0C0C0} 
\hpimap PO & 1.34         & 9.85        & Mixed                                                          & 23.083
          & 1.007 \\ \hline
\rowcolor[HTML]{C0C0C0} 
\hpimap PO+RR      & 2.25       & 10.39    & Mixed                                                          & \textbf{21.322} & \textbf{0.682}         \\ \hline
\end{tabular}
}
\label{tab:HomoMap}
\end{table}

\begin{table*}[htp]
\centering
\caption{Comparison of \hpimap's Pareto optimization and row-remapping strategy with homogeneous mapping solutions. While homogeneous mappings favor one metric at the expense of others, leading to high latency (PIM) or significant performance degradation (Photonics), \hpimap effectively balances latency, energy, and model performance. RR performance constraints are listed under each model's performance metric.}
\resizebox{\textwidth}{!}{
\begin{tabular}{c|ccc|ccc|ccc|ccc}
\hline
                             & \multicolumn{3}{c|}{\begin{tabular}[c]{@{}c@{}}Pythia-70M~|~TinyStories\\ Benchmark PPL 20.329\end{tabular}}                                                                          & \multicolumn{3}{c|}{\begin{tabular}[c]{@{}c@{}}Pythia-2.8B~|~TinyStories\\ Benchmark PPL 14.334\end{tabular}}                                                                            & \multicolumn{3}{c|}{\begin{tabular}[c]{@{}c@{}}MobileViT-S~|~Chest X-Ray\\ Benchmark Acc 0.9533\end{tabular}}                                                                       & \multicolumn{3}{c}{\begin{tabular}[c]{@{}c@{}}MobileViT-S~|~Military Assets\\ Benchmark Acc 0.8972\end{tabular}}                                                                   \\ \cline{2-13} 
\multirow{-2}{*}{Strategy}   & \multicolumn{1}{c|}{Latency(ms) $\downarrow$}     & \multicolumn{1}{c|}{Energy(mJ) $\downarrow$}       & \begin{tabular}[c]{@{}c@{}}Perplexity $\downarrow$ \\ (s.t. $\Delta$ 1.0)\end{tabular} & \multicolumn{1}{c|}{Latency(ms) $\downarrow$}      & \multicolumn{1}{c|}{Energy(mJ) $\downarrow$}        & \begin{tabular}[c]{@{}c@{}}Perplexity $\downarrow$ \\ (s.t. $\Delta$ 1.0)\end{tabular} & \multicolumn{1}{c|}{Latency(ms) $\downarrow$}       & \multicolumn{1}{c|}{Energy(mJ) $\downarrow$}      & \begin{tabular}[c]{@{}c@{}}Accuracy $\uparrow$ \\ (s.t. $\Delta$ 0.02)\end{tabular} & \multicolumn{1}{c|}{Latency(ms) $\downarrow$}      & \multicolumn{1}{c|}{Energy(mJ) $\downarrow$}      & \begin{tabular}[c]{@{}c@{}}Accuracy $\uparrow$ \\ (s.t. $\Delta$ 0.02)\end{tabular} \\ \hline
100\% SRAM                   & \multicolumn{1}{c|}{{\color[HTML]{CB0000} 10.21}} & \multicolumn{1}{c|}{{\color[HTML]{CB0000} 13.79}}  & 20.329                                                                                 & \multicolumn{1}{c|}{{\color[HTML]{CB0000} 411.83}} & \multicolumn{1}{c|}{{\color[HTML]{CB0000} 540.28}}  & 14.334                                                                                  & \multicolumn{1}{c|}{{\color[HTML]{CB0000} 291.92}}  & \multicolumn{1}{c|}{4.70}                         & 0.953                                                                               & \multicolumn{1}{c|}{{\color[HTML]{CB0000} 291.92}} & \multicolumn{1}{c|}{4.70}                         & 0.897                                                                               \\ \hline
100\% ReRAM                  & \multicolumn{1}{c|}{{\color[HTML]{CB0000} 14.73}} & \multicolumn{1}{c|}{{\color[HTML]{CB0000} 13.44}}  & 20.340                                                                                 & \multicolumn{1}{c|}{{\color[HTML]{CB0000} 581.62}} & \multicolumn{1}{c|}{{\color[HTML]{CB0000} 529.79}}  & 14.604                                                                                  & \multicolumn{1}{c|}{{\color[HTML]{CB0000} 583.54}}  & \multicolumn{1}{c|}{3.97}                         & 0.949                                                                               & \multicolumn{1}{c|}{{\color[HTML]{CB0000} 583.54}} & \multicolumn{1}{c|}{3.97}                         & 0.888                                                                               \\ \hline
100\% TeMPO~\cite{zhang2024tempo}                  & \multicolumn{1}{c|}{0.91}                         & \multicolumn{1}{c|}{8.92}                          & {\color[HTML]{CB0000} 23.839}                                                          & \multicolumn{1}{c|}{35.86}                         & \multicolumn{1}{c|}{361.92}                         & {\color[HTML]{CB0000} 16.419}                                                           & \multicolumn{1}{c|}{2.17}                           & \multicolumn{1}{c|}{{\color[HTML]{CB0000} 15.71}} & {\color[HTML]{CB0000} 0.883}                                                        & \multicolumn{1}{c|}{2.17}                          & \multicolumn{1}{c|}{{\color[HTML]{CB0000} 15.71}} & {\color[HTML]{CB0000} 0.827}                                                        \\ \hline
\rowcolor[HTML]{C0C0C0} 
\hpimap PO+RR & \multicolumn{1}{c|}{\cellcolor[HTML]{C0C0C0}2.25} & \multicolumn{1}{c|}{\cellcolor[HTML]{C0C0C0}10.39} & 21.322                                                                                 & \multicolumn{1}{c|}{\cellcolor[HTML]{C0C0C0}94.76} & \multicolumn{1}{c|}{\cellcolor[HTML]{C0C0C0}452.44} & 15.242                                                                                  & \multicolumn{1}{c|}{\cellcolor[HTML]{C0C0C0}117.34} & \multicolumn{1}{c|}{\cellcolor[HTML]{C0C0C0}6.19} & 0.942                                                                               & \multicolumn{1}{c|}{\cellcolor[HTML]{C0C0C0}99.46} & \multicolumn{1}{c|}{\cellcolor[HTML]{C0C0C0}8.25} & 0.883                                                                               \\ \hline
\end{tabular}
}

\label{tab:MainResults}
\end{table*}

\subsubsection{Heterogeneous vs. Homogeneous}
\begin{figure}
    \centering
    \includegraphics[width=0.99\columnwidth]{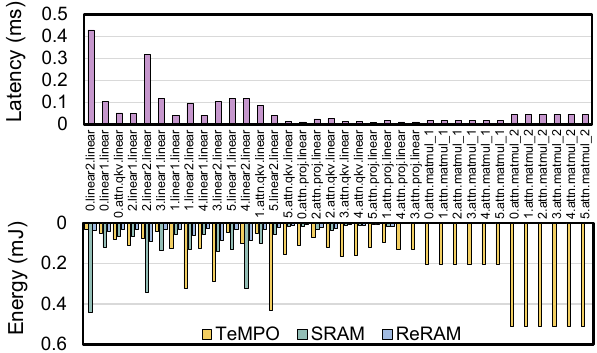}
    \caption{Layer energy and latency distribution of Pythia-70M after \hpimap Pareto optimization and row remapping. Layers are sorted from the most to the least sensitive one.}
    \vspace{-5pt}
    \label{fig:FinalDist}
\end{figure}

To validate our approach, we compare three methods: (1) our proposed framework, (2) an intuitive ``equal workload'' partition, and (3) a homogeneous workload assignment. Results are shown in Table~\ref{tab:HomoMap}. From the table, PO achieves superior efficiency: it reduces latency by 3.66\(\times\) and energy consumption by 1.22\(\times\) compared to the equal-workload method, and by 6.42\(\times\) on latency and 1.22\(\times\) on energy on average compared to the homogeneous assignment.
However, under the PO workload assignment, the model’s performance does not meet the 1.0-difference constraint; after RR, it improves substantially while still maintaining a good latency and energy profile.
For a comprehensive comparison, we introduce a Latency–Energy–Performance (LEP) score, defined as the sum of the normalized values of each metric group, where lower scores indicate better efficiency.
From Table~\ref{tab:HomoMap}, we see that PO pushes latency and energy close to the pure photonic mapping, and PO+RR then recovers accuracy to deliver the best overall LEP score.
Figure~\ref{fig:FinalDist} shows each layer's latency and energy after \hpimap optimization.
\subsubsection{\hpimap Effectiveness}
\begin{table}[]
\centering
\caption{Search time comparison between \hpimap and multi-objective search strategy performed on Pythia-70M.}
\begin{tabular}{|c|c|}
\hline
\multicolumn{2}{|c|}{\textbf{Optimizing Stop Threshold}}                                                                                             \\ \hline
\multicolumn{2}{|c|}{Latency $< 2.5$ ms, Energy $< 10.5$ mJ, $\Delta$ Perplexity $< 1.0$}                                                     \\ \hline
\multicolumn{1}{|c|}{\textbf{Strategy}}                                                                             & \textbf{Time Consuming (mins)} \\ \hline
\multicolumn{1}{|c|}{\begin{tabular}[c]{@{}c@{}}Multi-Object Search\\ (Latency + Energy + Perplexity)\end{tabular}} & 84.2                          \\ \hline
\multicolumn{1}{|c|}{\hpimap}                                                                                       & 51.3                           \\ \hline
\end{tabular}
\label{tab:timeconsume}
\end{table}

\begin{figure}
    \centering
    \includegraphics[width=0.7\columnwidth]{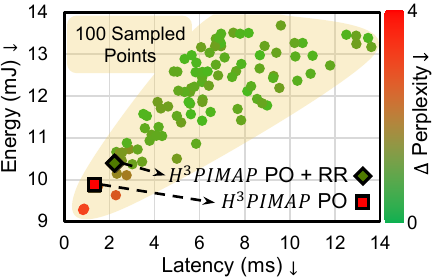}
    \caption{\hpimap optimized workload (Pythia-70M) mapping result compares with 100 randomly sampled mappings.}
    \label{fig:random_comp}
\end{figure}

To demonstrate \hpimap’s effectiveness, we measured optimization time against a full three‑objective evolutionary search, shown in Table~\ref{tab:timeconsume}.
A common termination criterion was imposed: latency < 2.5 ms, energy < 10.5 mJ, and $\Delta$perplexity < 1.0, slightly looser than \hpimap’s final solution to avoid bias.
Under this criterion, \hpimap converged in 51.3 minutes, whereas the pure multi‑objective search required 84.2 minutes, yielding a 39\% speed‑up.
We also benchmarked our mapping quality against 100 randomly sampled workload assignments, shown in Fig.~\ref{fig:random_comp}. 
The first‑stage Pareto Optimization (PO) drives latency and energy deep into the lower‑left corner of the design space; the second‑stage Row Remapping (RR) then recovers model fidelity with only a modest retreat in efficiency. 
The resulting PO + RR solution dominates most sampled points, achieving substantially lower energy and latency while still meeting the accuracy bound.
\subsubsection{Main Results}

Lastly, we compare the result of \hpimap mapping Pythia-70M (trained on TinyStories), Pythia-2.8B (fine-tuned on TinyStories), and MobileViT-S (trained on Chest X-Ray and Military Assets) to the proposed 3D heterogeneous architecture with homogeneous mapping, shown in Table~\ref{tab:MainResults}.

Homogeneous mappings (i.e., 100\% mapped to a single tier type) either incur large latency in pure PIM tiers or fail to meet the accuracy target on the pure photonic tier. Under matched-quality constraints (NLP models: $\Delta$PPL $\leq$ 1.0; vision models: $\Delta$Acc $\leq$ 0.02),
\hpimap delivers an average of 3.32$\times$ lower end-to-end latency versus the best valid homogeneous baseline across NLP and vision tasks. 
The advantage is strongest on LLMs: \textbf{\hpimap achieves 77.0$\%$ lower latency and 14.6$\%$ lower energy on Pythia-2.8B at marginal perplexity discrepancy}.

\section{Conclusion}
\label{sec:conclusion}
This work presents \hpimap, a heterogeneity-aware mapping framework that optimally distributes hybrid DNN workloads across a novel Electronic-Photonic-PIM heterogeneous AI computing architecture.
By integrating multi-objective optimization with accuracy-driven remapping, \hpimap achieves a Pareto-optimal balance of latency and energy efficiency while ensuring robust model inference against hardware non-ideal variations.
Across language and vision benchmarks, H3PIMAP delivers a 3.32$\times$ reduction in latency compared to homogeneous systems and naïve mappings; on LLMs, it further reduces latency by 77.0$\%$ and energy consumption by 14.6$\%$ under the constraint of matched task quality.
These results highlight the transformative potential of heterogeneous AI hardware, where the synergistic integration of electronic PIM and photonics, coupled with efficient software workload mapping, unlocks new frontiers in next-generation AI acceleration.

{\small
\balance
\bibliographystyle{ACM-Reference-Format}
\bibliography{ref/Top_sim,ref/Top, ref/addition,ref/Software,ref/NP,ref/ALG,ref/Cell,ref/PD,ref/DFM,ref/MPL,ref/NN, ref/IEEESettings}
}

\end{document}